\documentclass[conference]{IEEEtran}
\IEEEoverridecommandlockouts

\usepackage{cite}
\usepackage{amsmath,amssymb,amsfonts}
\usepackage{algorithmic}
\usepackage{graphicx}
\usepackage{textcomp}
\usepackage{xcolor}
\usepackage{subcaption}
\def\BibTeX{{\rm B\kern-.05em{\sc i\kern-.025em b}\kern-.08em
    T\kern-.1667em\lower.7ex\hbox{E}\kern-.125emX}}

\usepackage{booktabs}
\usepackage{caption}
\usepackage{textcase}
    
\begin{document}

\title{Understanding Communication Backends in Cross-Silo
Federated Learning
}

\author{%
\IEEEauthorblockN{%
Amir Ziashahabi\textsuperscript{\dag},
Chaoyang He\textsuperscript{\ddag},
Salman Avestimehr\textsuperscript{\dag}}
\IEEEauthorblockA{\textsuperscript{\dag}\textit{Department of Electrical and Computer Engineering}, University of Southern California, Los Angeles, CA, USA}
\IEEEauthorblockA{\textsuperscript{\ddag}\textit{TensorOpera, Inc., Palo Alto, CA, USA}
\IEEEauthorblockA{\texttt{ziashaha@usc.edu, ch@tensoropera.com, avestime@usc.edu}}}

}

\maketitle

\maketitle

\makeatletter
\def\ps@IEEEtitlepagestyle{
  \def\@oddfoot{\mycopyrightnotice}
  \def\@evenfoot{}
}
\def\mycopyrightnotice{
  {\footnotesize
  \begin{minipage}{\textwidth}
  \centering
  ~\copyright~2026 IEEE. Personal use of this material is permitted. Permission from IEEE must be obtained for all other uses, in any current or future media, including reprinting/republishing this material for advertising or promotional purposes, creating new collective works, for resale or redistribution to servers or lists, or reuse of any copyrighted component of this work in other works.
  \end{minipage}
  }
}
\makeatother

\begin{abstract}
Federated learning (FL) has emerged as a practical means for privacy-preserving distributed machine learning. FL's versatile design makes it suitable for various training settings, from IoT edge devices in cross-device FL to powerful servers in cross-silo FL. A key consequence of this versatility is the high level of diversity found in the networking configuration of FL applications. Coupled with the rising demand for large-scale models such as large language models, well-informed selection and configuration of communication backends become crucial for ensuring optimal performance in FL systems. This work focuses on cross-silo federated learning, presenting in-depth benchmarks of various communication backends, including MPI, gRPC, and PyTorch RPC. In addition, we introduce gRPC+S3, a hybrid backend designed to overcome the limitations of existing approaches, particularly for transmitting large models across geo-distributed deployments, achieving up to $3.8\times$ end-to-end speedup over gRPC. Our benchmarks examine point-to-point and end-to-end performance for a broad range of model sizes running under realistic network conditions. Our findings provide practical insights for selecting and configuring suitable communication backends tailored to the specific federated learning tasks and network configurations.
\end{abstract}



\begin{IEEEkeywords}
Federated learning, cross-silo, communication backends, MPI, gRPC, wide-area networks (WAN), cloud computing
\end{IEEEkeywords}

\section{Introduction}

Cross-silo federated learning (FL) is a distributed machine learning paradigm that enables collaboration among multiple organizations or data silos to train machine learning models while preserving data privacy \cite{kairouz2019advances}. A critical component in an FL system is its communication backend, which manages the exchange of model updates and control messages among participants. The performance of the communication backend has a profound impact on several key facets of the FL process, including convergence speed, communication overhead, fault tolerance, and resource utilization.


\begin{figure}
\begin{center}
  \includegraphics[width=1\linewidth]{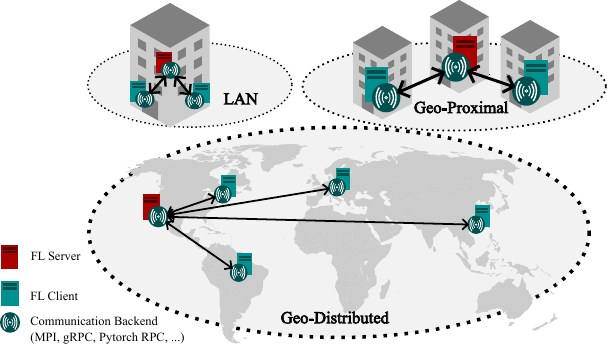}

  \caption{Three common deployment environments for federated learning: (1) LAN: participants in a single building; (2) Geo-Proximal: participants in close geographical proximity; (3) Geo-Distributed: participants in distant locations. Context-aware selection of communication backends is crucial for optimizing performance in each setting.
  }


\end{center}
\end{figure}

As the demand for FL systems grows and their applications expand, there is an increasing need to understand the performance implications and trade-offs associated with different communication backends, particularly in cross-silo FL scenarios, where options are numerous. This paper aims to provide an in-depth overview and performance comparison of various communication backends employed in cross-silo FL systems, including Message Passing Interface (MPI) \cite{forum1994mpi}, gRPC \cite{grpc}, and PyTorch RPC \cite{damania2023pytorch}. Additionally, in response to the limitations of these communication backends in the context of FL systems, we propose a novel communication backend that combines the capabilities of gRPC and Amazon Simple Storage Service (S3) \cite{aws_2018}. This hybrid approach aims to leverage the strengths of both gRPC and Amazon S3 to deliver enhanced performance, reliability, and scalability, particularly in situations where existing communication backends exhibit shortcomings.

We study communication backends for cross-silo FL through point-to-point and end-to-end experiments that vary message size and concurrency across three deployment environments: LAN, Geo-Proximal, and Geo-Distributed. In summary, this work offers the following contributions:

\begin{enumerate}
\item \textbf{Introduction of Hybrid gRPC+S3 backend.} A design that routes metadata over RPC and model weights via object storage. It is particularly useful for scenarios that involve transmitting large models across a geo-distributed network.
\item \textbf{Benchmarking suite across regimes.} 
A comprehensive benchmark suite that includes a variety of models, metrics, and communication backends tailored specifically for cross-silo FL settings. This enables researchers and practitioners to evaluate the performance of different backends across a diverse set of scenarios.

\item \textbf{Realistic deployment footing.} Experiments are conducted in realistic settings that are representative of cross-silo FL environments. We utilize GPU-powered machines and consider various training environments to ensure our results are applicable and insightful to real-world FL tasks.
\end{enumerate}



\section{Background and Motivation}

\subsection{Federated Learning}
Federated learning (FL) trains a shared model across decentralized clients without revealing raw data \cite{kairouz2019advances,pmlr-v54-mcmahan17a,konevcny2016federated}. There are two primary categorizations of FL: \emph{cross-device} (many resource-constrained clients, e.g., phones/IoT) and \emph{cross-silo} (fewer but well-provisioned participants such as hospitals, banks, or data centers). A server coordinates rounds of training by selecting clients, distributing the current global model, collecting locally computed updates (e.g., via SGD \cite{robbins1951stochastic}), and aggregating them. This work focuses on cross-silo FL.

\subsection{Communication Backends for Distributed ML}\label{ssec:dml-comm-backends}
Communication backends move models, gradients, and control messages. Hence, the backend choice strongly affects runtime, scalability, and robustness. We study three widely used options: \textbf{MPI} \cite{forum1994mpi} (HPC-standard message passing with efficient point-to-point and collectives), \textbf{gRPC} \cite{grpc} (HTTP/2-based RPC with Protobuf and support for streaming, and TLS), and \textbf{PyTorch RPC} \cite{damania2023pytorch} (PyTorch framework-integrated RPC supporting sync/async patterns). 

\subsection{Motivation}

\begin{figure}[t]
    \centering
    \begin{minipage}[b]{0.42\textwidth}
        \centering
        \resizebox{\columnwidth}{!}{%
        \begin{tabular}{llccc}
        \toprule
        \multicolumn{2}{c}{\textbf{Region}} & \multicolumn{2}{c}{\textbf{Bandwidth (MB/s)}} & \textbf{Latency (ms)} \\
         & & \textbf{Single} & \textbf{Multi} & \\ \midrule
          & North California  & 592   & 2946 & 0.44 \\
          & Oregon            & 133   & 573  & 11 \\
          & North Virginia    & 39.4  & 557  & 32.3 \\
          & Hong Kong         & 16.3  & 513  & 83.3 \\
          & Stockholm         & 11.4  & 495  & 90.9 \\
          & Sao Paulo         & 8.27  & 491  & 90.9 \\
          & Bahrain           & 6.90  & 444  & 111 \\
        \bottomrule
        \end{tabular}}
        \captionof{table}{EC2 bandwidth (single vs.\ multi-connection) and latency between North California and other regions (g4dn.2xlarge).}
        \label{tab:bandwidth}
    \end{minipage}
    \hfill
    \begin{minipage}[b]{0.42\textwidth}
        \centering
        \includegraphics[width=\linewidth]{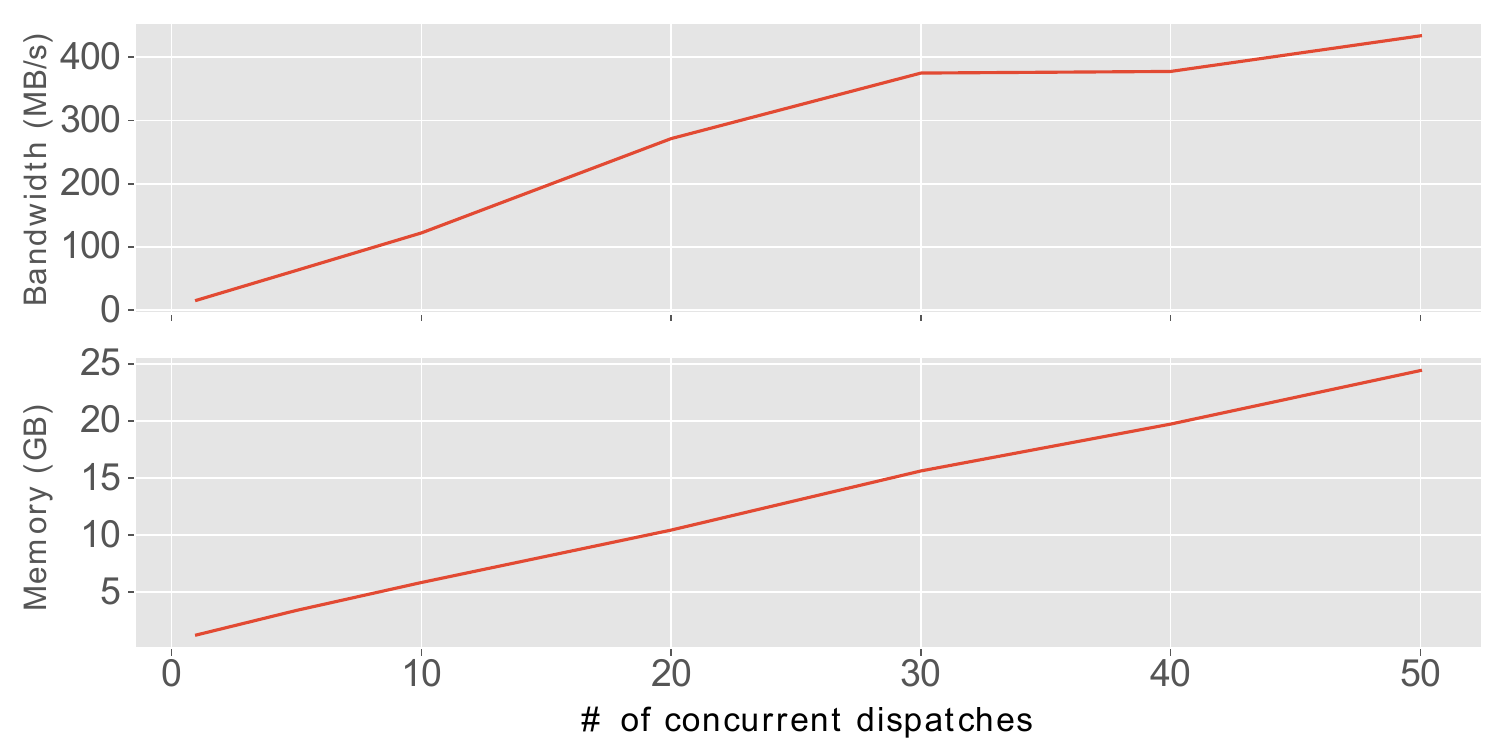}
        \caption{Effect of concurrent dispatch on gRPC: bandwidth (top) and memory (bottom) for North California $\rightarrow$ Bahrain.}
        \label{fig:mem_bandwidth}
    \end{minipage}
\end{figure}

\textit{(1) FL’s diverse training environments and lack of benchmark studies.} Cross-silo FL runs over vastly different networks, from fast intra-site fabrics to high-latency wide area networks (WANs), yet systematic evaluations of communication backends across these regimes are scarce. We quantify this variability on two platforms: a LAN with InfiniBand (up to 5~GB/s bandwidth and 3.17~$\mu$s latency; TCP fallback about 1~GB/s and 16.8~$\mu$s) and AWS EC2 with g4dn.2xlarge instances. Table~\ref{tab:bandwidth} reports single- and multi-connection bandwidth and latency between North California and seven regions: intra-region links reach $\sim$3~GB/s and 0.44~ms, while inter-region links are much slower. The gap between single- and multi-connection throughput widens as latency increases, making connection-level concurrency a first-order factor. This work addresses the benchmarking gap by evaluating backends under LAN, geo-proximal, and geo-distributed conditions with both microbenchmarks and end-to-end training.

\textit{(2) FL network characteristics and limitations of existing backends.} Cross-silo FL deployments require backends that (i) perform under heterogeneous networks with minimal assumptions, (ii) tolerate dynamic participation and failures without global aborts, (iii) scale in both client count and model size while keeping CPU/memory bounded, and (iv) support secure operation over untrusted WAN. Given these needs, we revisit the backends in \S\ref{ssec:dml-comm-backends} for FL. \emph{MPI}, while efficient, assumes static membership and homogeneous, tightly managed networks. Moreover, SSH/rsh-based process management complicates deployment across independently administered sites, and failure handling often lacks the fault isolation needed when participants can come and go. \emph{PyTorch RPC} integrates cleanly with the framework but is framework-specific, expects open and stable paths between peers, and offers less control over transport/topology than MPI/gRPC, which can be restrictive under stricter cross-organization security and networking constraints. \emph{gRPC} is a popular choice for FL applications due to its flexibility and security features. However, it is not optimized for heavy data transfers. For instance, its standard Python implementation multiplexes all traffic over a single HTTP/2 connection. Where multi-connection throughput greatly exceeds single-connection (Table~\ref{tab:bandwidth}), a single channel underutilizes the link. Using many channels can recover throughput for large broadcasts yet raises server memory roughly linearly with concurrency because each send buffers its own copy (Fig.~\ref{fig:mem_bandwidth}). Motivated by these constraints, especially for large message transfers across heterogeneous, untrusted networks, we propose gRPC+S3, a communication backend optimized for federated learning scenarios.


\section{\textnormal{g}RPC+S3: A Hybrid Communication Backend for Federated Learning}

To address the limitations of common communication backends in the context of FL systems, we propose a novel communication backend named gRPC+S3. This approach combines the flexibility and security benefits of gRPC with the highly scalable and reliable Amazon S3 storage to provide an efficient and FL-friendly communication method.

\subsection{Architecture}
\begin{figure}[t]
\begin{center}
\includegraphics[width=1\linewidth]{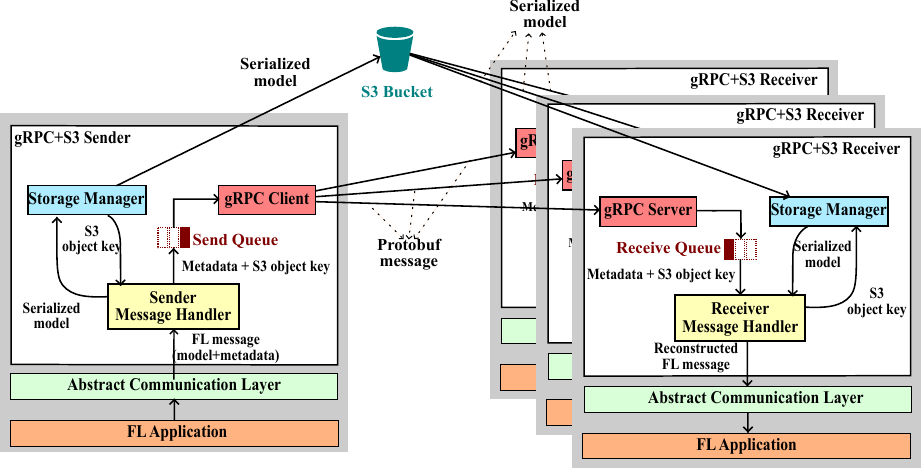}
\caption{Architecture of the proposed gRPC+S3 backend.}
\label{fig:grpc_s3_arch}
\end{center}
\end{figure}

Each FL message can be split into a small metadata record (round, type, sender ID, object key) and a large parameter payload (the model). Transfer using gRPC+S3 is done in two phases. \emph{Sender side:} (1) upon a send, the Sender Message Handler separates metadata from the model; (2) if the model is new, it is serialized and uploaded to an S3 bucket via the Storage Manager, which returns the S3 object key; repeated sends of the same model reuse a cached key; (3) the handler constructs a compact Protobuf containing metadata and the object key and enqueues it for the gRPC client, which delivers it to the receiver’s gRPC server. \emph{Receiver side:} (1) the gRPC server enqueues incoming records; (2) the Receiver Message Handler extracts the object key and fetches the serialized model via the Storage Manager; (3) the model is deserialized and recombined with metadata to reconstruct the original FL message.

This split keeps RPC messages small and shifts the heavy data path to S3, where receivers can pull in parallel using independent connections.

\subsection{Advantages}\label{ssec:grpc_s3_adv}
\textit{Efficiency.} Large parameters bypass the single-connection constraint typical of a Python gRPC channel and leverage S3’s multi-connection downloads; gRPC carries only lightweight control, reducing per-transfer CPU and copying on the sender.

\textit{Scalability.} Broadcasts become single-upload/multi-download: the server uploads once, and all clients fetch concurrently. The server’s peak memory during broadcast is independent of the number of receivers because it does not need to buffer $N$ concurrent copies.


\textit{Versatility.} Deployments can fall back to pure gRPC when the payload is small, the environment lacks object storage, or S3 round-trips would dominate.

\textit{Reliability and fault tolerance.} gRPC provides robust request handling and retries for metadata. S3 offers durable object storage and independent client retrieval, so late or failed receivers can re-fetch without forcing the sender to re-transmit.

\textit{Privacy and security.} Both legs support encryption in transit (TLS for gRPC; HTTPS for S3). Access to model objects can be controlled via scoped credentials or time-limited, pre-signed URLs.


\section{Benchmarking Methodology}
We analyze communication backend performance across common cross-silo FL scenarios. We report peer-to-peer and end-to-end results over diverse models and deployment environments. Below we summarize the environments and workloads used in our benchmarks.

\subsection{Environments}\label{ssec:env}
We evaluate three deployment regimes that frequently arise in cross-silo FL.

\textbf{Local Area Network (LAN).}
This setting represents a high-bandwidth, low-latency environment typical of a single data center. Our testbed consists of two machines, each with 8 NVIDIA Quadro RTX 5000 GPUs, connected via a 5 GB/s InfiniBand link.

\textbf{Geo-Proximal.}
Participants are in nearby sites (e.g., same metro/region) with moderate latency and high available bandwidth.We emulate it using Amazon EC2 g4dn.2xlarge instances in different availability zones within the North California region. This setup provides high-speed connectivity, with measured bandwidth of 592 MB/s for single-threaded and 2946 MB/s for multi-threaded communication.

\textbf{Geo-Distributed.}
 This simulates collaboration across continents over a high-latency WAN. We use Amazon EC2 g4dn.2xlarge instances, placing the server in North California and clients in seven other regions: North California, Oregon, North Virginia, Hong Kong, Stockholm, Sao Paulo, and Bahrain. Please see Table~\ref{tab:bandwidth} for the detailed network characteristics of this environment.

\subsection{Models and Datasets}
We evaluate both image and text-based tasks to cover common FL workloads, using CNNs and Transformers across four payload-size tiers. Image tasks use GLD-23K \cite{weyand2020google}; text tasks use 20 Newsgroups \cite{lang1995newsweeder}. The tiers are: Small (ResNet56 \cite{he2016deep}, 591{,}322 parameters, 2.39 MB), Medium (MobileNetV3 \cite{howard2019searching}, 5{,}152{,}518 parameters, 19.85 MB), Big (DistilBERT \cite{sanh2019distilbert}, 66{,}362{,}880 parameters, 253.19 MB), and Large (ViT-Large \cite{dosovitskiy2020image}, 307{,}432{,}234 parameters, 1{,}243.14 MB).

These tiers are representative of FL payloads in practice. Even when targeting Large Language Models (LLMs) with billions of parameters, practical deployments typically reduce communication with parameter-efficient fine-tuning techniques (e.g., LoRA) \cite{hu2022lora, babakniya2023slora}, so the transferred state is closer to these tiers than to full model checkpoints.

\begin{figure*}[t]
  \centering
  \includegraphics[width=0.9\textwidth]{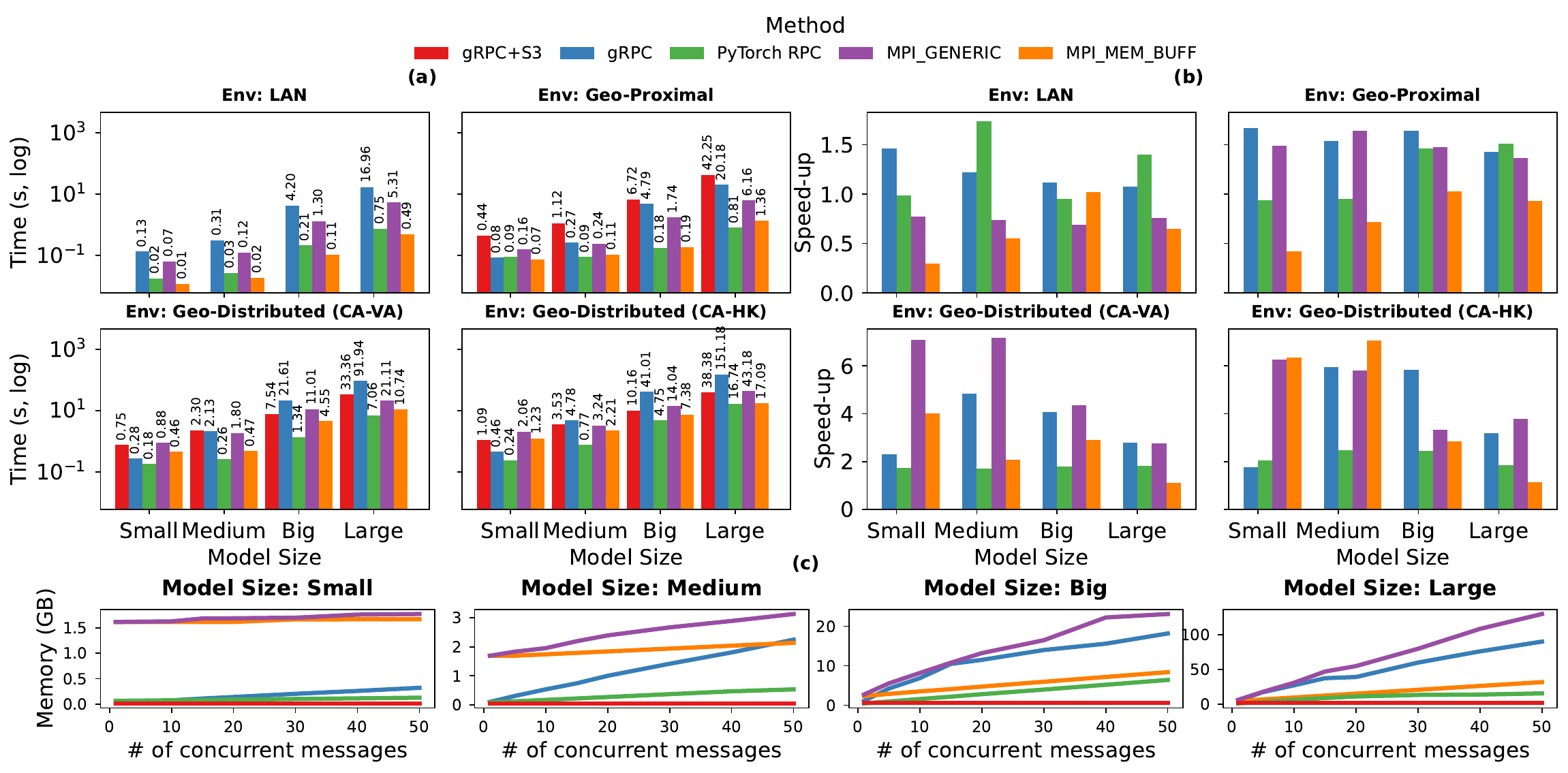}
  \caption{Peer-to-peer results across backends, environments, and model sizes.
(a) CPU-to-CPU latency (log scale).
(b) Speedup of concurrent over sequential transmission for 10 messages (Large uses 5).
(c) Peak sender memory during concurrent broadcast.}
  \label{fig:p2p}
\end{figure*}

\subsection{Peer-to-Peer Benchmarks}
We measure backends in isolation to examine how message size and environment shape transport behavior. Metrics include latency (CPU-to-CPU), impact of concurrent sends, and memory utilization. Backends are \texttt{gRPC}, \texttt{PyTorch RPC}, \texttt{gRPC+S3}, and two MPI variants: \texttt{MPI\_GENERIC}, which serializes and transforms arbitrary Python objects,  and \texttt{MPI\_MEM\_BUFF}, which uses pre-allocated buffers and achieves near C-level speeds but can only communicate buffer-like objects. 
To capture the variability of Geo-Distributed deployments, we measure two scenarios: (1) intra-continent communication between North California and North Virginia (CA–VA) and (2) inter-continent communication between North California and Hong Kong (CA–HK).  
\texttt{gRPC+S3} is excluded from LAN experiments, since its reliance on external object storage would make S3 access latency the dominant factor, masking backend performance.

\subsection{End-to-End Benchmarks}\label{ssec:e2e-benchmark}
For realistic behavior, we take the best configuration per backend, integrate it into FedML \cite{he2020fedml}, and run full FL training for each model tier. We record wall-clock time for communication, CPU–GPU migration, serialization, and waiting on both clients and the server. Client measurements also include training time, while the server adds aggregation time. Waiting corresponds to idle or blocked states, and serialization applies only to \texttt{gRPC} and \texttt{gRPC+S3}. We report client times averaged across all clients.
As above, \texttt{gRPC+S3} is omitted on LAN.


\subsection{Implementation and Configuration Details}\label{ssec:implementation}
\textbf{PyTorch RPC.} We use the Distributed RPC Framework \cite{damania2023pytorch} with TensorPipe. CUDA RPC is enabled (GPU-to-GPU transfers via device maps) for suitable end-to-end experiments. 
To make it compatible with multi-region AWS configuration, nodes were placed in VPCs with pairwise peering.

\textbf{gRPC / gRPC+S3.} \texttt{gRPC} uses \texttt{grpcio} \cite{grpc}. Unary and Streaming modes performed similarly in peer-to-peer tests, so results apply to either. \texttt{gRPC+S3} uses Streaming RPC for metadata/control and \texttt{boto3} for S3.

\textbf{MPI.} We use CUDA-aware Open MPI \cite{gabriel2004open} over UCX \cite{shamis2015ucx} using \texttt{mpi4py} \cite{dalcin2021mpi4py} interface. \texttt{MPI\_GENERIC} uses lowercase calls (e.g., send) to serialize arbitrary Python objects. \texttt{MPI\_MEM\_BUFF} uses uppercase calls (e.g., Send) to transfer buffer-like objects.


\begin{figure*}[t]
  \centering
  \includegraphics[width=0.9\textwidth]{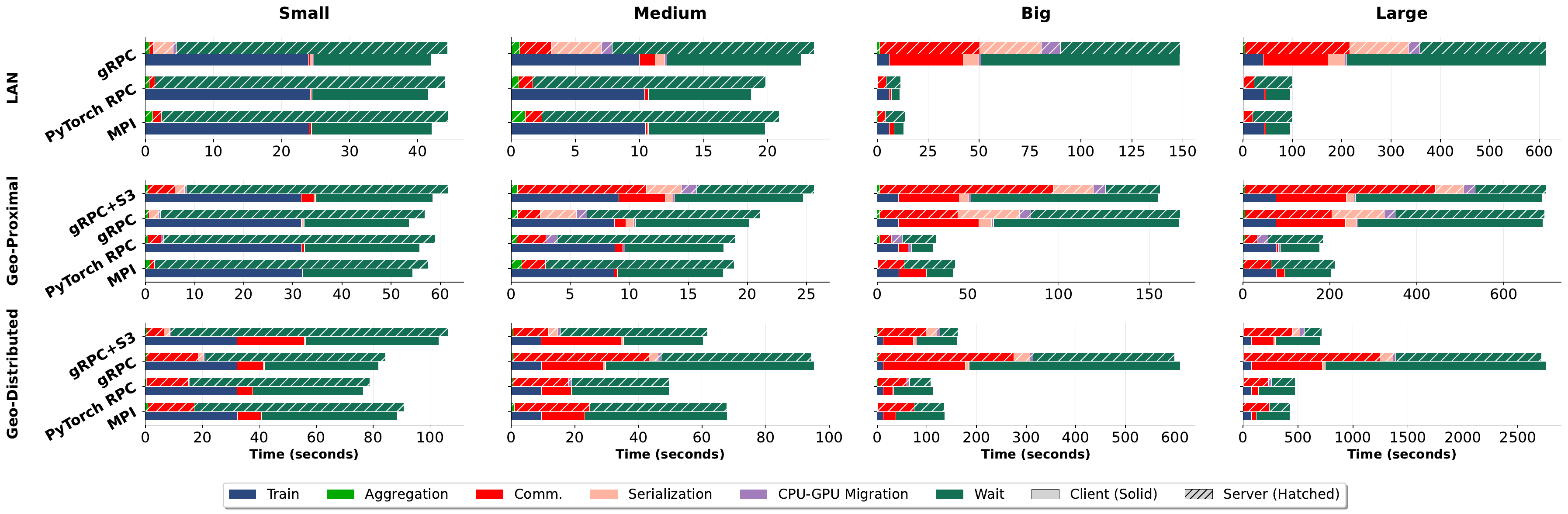}
  \caption{Per-state duration in end-to-end LAN experiments (communication, CPU-GPU migration, serialization, waiting; plus training on clients and aggregation on server).}
  \label{fig:e2e}
\end{figure*}
\section{Peer-to-Peer Experiment Results}

As shown in Figure~\ref{fig:p2p}(a), the optimal communication backend is highly dependent on the network environment. In low-latency LAN and Geo-Proximal settings, backends that avoid serialization overhead, such as \texttt{MPI\_MEM\_BUFF} and \texttt{PyTorch RPC}, consistently deliver the best performance. The impact of serialization is significant; for instance, in the LAN environment, serialization accounts for up to 86\% of \texttt{gRPC}'s total latency and is the primary reason for \texttt{MPI\_GENERIC}'s inferior performance compared to \texttt{MPI\_MEM\_BUFF}. As we move to high-latency Geo-Distributed environments, proficiency in utilizing multiple connections becomes the dominant factor for performance. Here, \texttt{PyTorch RPC} consistently outperforms other backends across most model sizes. In contrast, \texttt{gRPC}'s lack of a built-in mechanism for multi-connection transfers causes its performance to degrade substantially with larger models in this setting.

As illustrated in Figure~\ref{fig:p2p}(b), concurrent message transmission, a common scenario for an FL server distributing a global model, generally enhances performance. This is particularly true in Geo-Distributed settings where it mitigates the effect of high network latency and improves bandwidth utilization, leading to substantial speedups for all backends. \texttt{gRPC} and \texttt{MPI\_MEM\_BUFF} see performance gains of up to 7$\times$ in these conditions. However, there is an exception in the LAN environment, where MPI-based backends experience a performance decline with concurrent transmissions, likely due to multithreading overheads offsetting communication gains.

Memory utilization during concurrent dispatches, detailed in Figure~\ref{fig:p2p}(c), reveals stark differences between backends. \texttt{gRPC+S3} is exceptionally memory-efficient, as it uploads the model to S3 only once and subsequently sends lightweight metadata messages, resulting in constant memory usage regardless of the number of receivers. In contrast, memory consumption for all other backends increases almost linearly with the number of concurrent dispatches. This issue is most severe for \texttt{gRPC} and \texttt{MPI\_GENERIC}, which create a separate copy of the data for each message, making concurrent dispatches of large models impractical. \texttt{MPI\_MEM\_BUFF} and \texttt{PyTorch RPC} are more memory-efficient as their design allows them to read directly from buffer memory, avoiding data duplication.

\section{End-to-End Experiments}

In our end-to-end experiments, we evaluated the system performance using one server and seven clients, with clients training for only a single epoch in each round to emphasize the impact of communication backends. The server was configured to use concurrent transmission to distribute the global model, and each backend was set to its optimal configuration (e.g., GPU-Aware MPI and environment-specific settings for PyTorch RPC).

The results are shown in Figure \ref{fig:e2e}. In  LAN and Geo-Proximal environments, the total execution time for small and medium-sized models is heavily dominated by the client training time, leading to comparable performance across most backends. The primary exception is \texttt{gRPC+S3}, whose two-step transfer mechanism introduces overhead that makes it less competitive in these settings. As model sizes increase, communication becomes the bottleneck, and the efficiency of \texttt{MPI} and \texttt{PyTorch RPC} allows them to significantly outperform other backends. In the LAN setting, \texttt{gRPC}-based training was approximately 9$\times$ slower for large models.

In the Geo-Distributed environment, communication latency is the primary performance driver, leading to high variance in client completion times, resulting in long waiting periods. \texttt{PyTorch RPC} delivered the lowest execution time for most model sizes, with MPI performing closely and even surpassing it for large models. The most notable result in this setting is the performance inversion between \texttt{gRPC} and \texttt{gRPC+S3}. While \texttt{gRPC} is competitive for small models, it struggles with larger ones because it utilizes available bandwidth ineffectively. This limitation is overcome by \texttt{gRPC+S3}, which delegates the large data transfer to S3, making it \textbf{3.5--3.8$\times$ faster than \texttt{gRPC}} for large models and a much more suitable choice for geo-distributed deployments.

\section{Discussion}

The optimal communication backend hinges on a trade-off between the network environment and practical deployment requirements. In low-latency trusted networks (LAN, Geo-Proximal), performance is limited by CPU overheads like serialization, making memory-buffer backends (\texttt{MPI}, \texttt{PyTorch RPC}) superior. Conversely, in high-latency Geo-Distributed settings, efficient connection concurrency is paramount, giving \texttt{PyTorch RPC} a general advantage. The high-latency context also reveals a critical bottleneck in standard \texttt{gRPC} for large models, a limitation our \texttt{gRPC+S3} backend resolves by offloading transfers to cloud storage, improving both throughput and memory efficiency. Practically, while \texttt{MPI} and \texttt{PyTorch RPC} are fastest, their unsuitability for untrusted WANs makes the versatile \texttt{gRPC+S3}, with \texttt{gRPC} fallback for small payloads ($<10$ MB), the most robust choice for many real-world FL deployments.

\section{Related Work}

Numerous federated learning (FL) frameworks have been proposed to benchmark tasks and models \cite{he2020fedml, beutel2020flower, lai2022fedscale}. These studies, however, typically focus on computational efficiency and model accuracy while overlooking the impact of the communication layer. Most benchmarks rely on a single default backend (usually \texttt{gRPC}) and lack a comparative analysis of alternative transport mechanisms, a gap our work directly addresses.

Our research also connects to two broader areas in distributed machine learning (DML). The first is work on characterizing and enhancing communication backends \cite{xu2022arm, damania2023pytorch}. These studies are generally confined to high-performance, trusted networks, whereas our analysis extends to untrusted, geo-distributed settings. The second area involves algorithmic techniques to reduce communication overhead, such as quantization \cite{alistarh2017qsgd} and sparsification \cite{wangni2018gradient}. These methods are orthogonal to our study and can be used in conjunction with any backend. Similarly, research in geo-distributed training has focused on algorithmic improvements like finding optimal topologies \cite{marfoq2020throughput} rather than providing a system-level analysis of the underlying communication backends, which is the focus of our paper.

\section{Conclusion}
In this study, we analyzed various communication backends (MPI, gRPC, and PyTorch RPC) for cross-silo FL. We employed peer-to-peer and end-to-end benchmarks to examine their performance under different network conditions and with varying model sizes. We also introduced a novel communication backend, gRPC+S3, to overcome the shortcomings of existing backends, particularly when transmitting large models in untrusted environments. In the context of federated learning, gRPC+S3 demonstrated remarkable memory efficiency, and our Geo-Distributed experiments show that gRPC+S3 significantly outperforms gRPC for messages larger than 10 MB (up to $3.8\times$), making it a practical choice in such settings. Finally, this study offers clear guidelines and insights, helping researchers and practitioners choose and configure the appropriate communication tools for their federated learning tasks to build efficient systems.

\section{Acknowledgment}
This work was supported in part by Qualcomm Technologies, Inc. The opinions, findings, and conclusions or recommendations expressed are those of the author(s) and do not necessarily reflect the views of the sponsor.

\bibliographystyle{IEEEtran} 
\bibliography{ref} 

\end{document}